# Self-Assembly in the Growth of Precious Opal

A. M. Stewart[a], Lewis T. Chadderton[b*] and Brian R. Senior[c]

[a] *Emeritus Faculty, The Australian National University, Canberra, ACT 0200, Australia*
[b] *Gonville and Caius College, Trinity Street, Cambridge CB2 1TA, United Kingdom*
[c] *Senior & Associates, 303 Shingle Hill Way, Bungendore, NSW 2621, Australia*
[*] *On leave from The Atomic and Molecular Physics Laboratories, School of Physics and Engineering, The Australian National University, Canberra, ACT 0200, Australia.*

**Abstract**
It is proposed that primary nucleation of amorphous microspherulites of hydrated silica in natural proto-precious-opal can be followed by a long range superlattice ordering process by means of electrostatic self-assembly. Necessary conditions in the thermodynamics are a high surface charge density on microspherulite surfaces, a long Debye length and an appropriate number density of nucleation centres. A further chemical requirement is a high alkaline environmental $pH \sim 9$–$10$. It is also proposed that the characteristic concentric spherical shell-like structure of spherulites, centred on primary nuclei, are due to sequential deposition of intrinsic salts which precipitate out when the corresponding solubility limits in the liquid are successively exceeded. It can be that the better-known sedimentation of microspherulites under gravity only plays part in the final stabilization period of overall growth.

1.     **Introduction**

It is well known that microspherulites of amorphous hydrated silica ($SiO_2 \approx 27H_2O$) in natural opal form a characteristic superlattice [1] with lattice translation parameters in the range ~150 - 450 nm (Fig. 1), and that it is simple Bragg diffraction of light from such nanostructures that gives precious opal its singular display of colour or 'fire' [2-3]. A simple model for the slow and prolonged growth of the individual spherulites in the natural geological environment, hereafter referred to as I, has recently been proposed [4]. Structural characteristics of a single spherulite strongly suggest that there is nucleation on a central core which, it is conjectured in I, carries the heavy common salts of uranium, thorium and lead. These specific compounds are identified by strong signals in SIMS (secondary ion mass-spectroscopy) analyses in precious opal. A final proof of this hypothesis, however, probably lies in part in successful fission track mapping experiments of natural stones. Yet this is neither a trivial nor simple experiment when concentrations of active compounds are only a few parts per million, which suggests that the more exceptional uraniferous opal might be a more quickly rewarding target. For common opal 'potch', on the other hand, in which central nuclei are most frequently absent (see I), the corresponding SIMS signals are very low. It is also the case that when clear central nuclei are present there are, in addition, up to three concentric sharp spheres that are centred upon individual nuclei. These, and the overall superlattice structures, are clearly and consistently seen in TEM micrographs of shadowed extraction replicas, and in SEM images from polished surfaces.

There is an anomalously high gamma radioactivity measured in drill holes in the vicinity of opal deposits in the field. It is very likely that this has its overall origin in natural spontaneous fission of the cited isotopes of thorium and uranium, and from their corresponding daughter product decay chains. And certainly there is a clear observed association of natural radioactivity with the abundance of precious opal which led directly to the extraordinary successful new exploration method for this important mineral described in I.

In a more macroscopic and geological picture it is found that amorphous silica cement is very abundant in deeply weathered profiles in which precious opals are found. This evidently appears to have been derived from the breakdown of former litho-feldspathic sedimentary rocks and smectitic clays, into kaolinite. And this process took place in the late Oligocene to early Miocene periods during a deep weathering event which defined the Canaway profile in Australia and was accompanied by strong silicification episodes forming porcellanite and silcrete [5]. Opals formed in the uppermost 35 m of this profile over times of up to ~ 20 My, and at various depths that accord with porosity and permeability barriers located at sandstone/claystone interfaces. Fractures within the profile acted as conduits for movement of silica-laden water and stress-related dilation zones and cavities formed by leaching of former soluble minerals or organic fragments, formed spaces in which siliceous water accumulated [6-8].

The detailed specific physics and chemistry of siliceous groundwater, the nucleation of individual spherulites with internal structure, and the corresponding long term sol-to-gel transformations, which lead to formation of both stable precious opal and potch, and which appear to take place simultaneously in a changing physical-chemical environment, are still poorly understood. It is generally assumed that it is sedimentation of growing spherulites under gravity which is responsible for ordering of the long-range superlattice. In this paper we critically consider nucleation and growth anew, present an explanation for the origin of the concentric shell structure in spherulites and, through thermodynamics, successfully describe a long-range interaction between them, which can be the chemico-physical and thermodynamic origin of the intrinsic precious opal superlattice (Fig. 1).

## 2. Mechanism

There exists a detailed literature on the growth of both precious and artificial opal [9-19] in which processes of self-assembly similar to those which operate for $C_{60}$ (fullerene) and corresponding nanotubes, or other quasi one- dimensional structures, are qualitatively considered. In general these approaches are based upon earlier studies on the fundamentals of colloid crystallization [20-27] and we have found no satisfactory physical outline of the critical processes governing nucleation and what immediately follows. A useful overview of the demands asked of the basic physics can be found in Norris *et al.* [9] who hypothesize that the initial ordering process is in some way due to solvent flow and that thereafter, as in many publications, it is sedimentation under gravity which is the dominating feature of growth.

Our own quite different model for the growth of natural [1, 2] and artificial [28] opal is based on electrostatic self-assembly and is, at this stage, essentially semi-quantitative at best. The system we consider is that of a silica-water metallic salt fluid comprising, for present purposes, sodium chloride (monovalent ions) or barium sulphate (divalent ions), and the ionic silicate species. When either the temperature is lowered or water is lost by evaporation or diffusion through the surrounding rock, such a solution tends to solidify into a solid hydrated silicate [29]. We adopt heterogeneous nucleation as the base mechanism from chemical reaction rate theory [30-32], and assume that the nucleation centres are situated randomly in the solution. The many body aggregation process is then closely akin to that described for self-assembly of the $C_{60}$ fullerene molecule [33-34] - either homogeneously or heterogeneously. As soon as thermodynamic conditions become favourable, growth of spherical solid objects (microspherulites of hydrated silica) starts simultaneously at these pre-existing randomly positioned nucleation centres and proceeds according to the standard mechanisms of crystal growth in supersaturated solutions. Our model invokes electrostatic interaction between the microspherulites.

A colloidal system of this type–with an electric double layer–has been earlier described in the theory of Derjaguin, Landau, Verwey and Overbeek (DLVO) [35]. Because of the discontinuity of the chemical and electrical properties at the solid-liquid boundary, the surfaces of the microspherulites acquires an electric charge density $\sigma$, due to the charge dissociation of silanol groups [29]. This charge is compensated by an oppositely charged Debye cloud in the electrolyte close to the surface which falls away exponentially with distance and establishes the double layer. Each microspherulite has a surface charge:

$$Q = 4\pi R^2 \sigma \tag{1}$$

where $R$ is the radius of the microspherulite. Accordingly, there is a repulsive screened Coulomb force between any two microspherulites (Fig. 2) separated by a distance $D$ ($D > 2R$) giving rise to electrostatic interaction energy $U$. We take this to be of the qualitative form [29, 37]:

$$U = \frac{Q^2}{4\pi\varepsilon_0\varepsilon_r} \frac{e^{-(D-2R)/L_D}}{D} \tag{2}$$

giving:

$$U = \frac{4\pi R^4 \sigma^2}{\varepsilon_0\varepsilon_r} \frac{e^{-(D-2R)/L_D}}{D} \tag{3}$$

where $L_D$ is the Debye length. For water of laboratory purity the Debye length is not greater than around 150 nm [35, 41]. The greater impurity level of natural waters is expected to lead to Debye lengths significantly less, so that the system will be in the regime $L_D < D$, as shown in Fig. 2. If $U$ is much less than $kT$ we have a system of solid microspherulites randomly positioned in the liquid, the influence of the interaction being swamped by thermal fluctuations. Conversely, however, if $U$ is much greater than $kT$ then it will be energetically favourable for the system to order into a regular array of small solid microspherulites situated within the liquid. This is of course, by definition, a many body problem. For no Debye screening ($L_D \to$ infinity) calculations show that the structure with minimum energy is fcc [36], with a neighbouring hcp structure with only very slightly greater energy. An fcc structure with stacking faults is therefore to be expected (Fig. 1). As the radius of the microspherulites grows, the prefactor of $U/kT$ ($4\pi\sigma^2 R^4 / D\varepsilon_0\varepsilon_r kT$) rises steeply as $R^4$, and the interaction energy increases rapidly as the Debye shells overlap.

In addition to calculating the effects of the fundamental many-body physics there is the requirement to calculate accurately the energy of interaction between the microspherulites for a given separation distance. This can only be achieved through complete solution of the full non-linear Poisson-Boltzmann equation including charge regulation and the effects of the finite radius of curvature of the microspherulites [35, 37]. Such a thorough and complex analysis is beyond the scope of this paper, which means that the results we do present cannot be taken to be more than an order-of-magnitude estimate. However, because the system is macroscopic, a sharp phase change is likely to take place from a positionally disordered to an ordered phase as the tendency to crystallisation increases by cooling and/or loss of water.

Growth of the solid microspherulites continues after their positions transform into a regular array, and all growth is arrested through exhaustion of the solvent when they make neighbour contact. At this point they become bound together by attractive van der Waals forces [38-39] into a true molecular solid. It is clear that it is easier to form a periodic array in the liquid than in the solid. Yet it is to be remarked upon that sedimentation under gravity, although possibly helpful in forming a consolidated structure, is neither a necessary nor a major feature of this model. We note that, because during the later stages of growth the radius of the spherulites will be proportional to the square root of the time from genesis (see Appendix A), there will be a tendency to narrow the relative size distribution of the spherulites as the later starters will be able to catch up with those nucleated earlier.

It is normal to suppose that because of the attractive van der Waals force the true ground state of any colloidal system is always the coagulated one, the colloidal state being at best metastable [35]. For present purposes we set this general notion aside and accept that a regular structure may be favoured thermodynamically, and that kinematic factors will be of importance in determining whether the microspherulites transform into a regular structure, giving precious opal, or quench into a random array, giving potch. In contrast to simpler glassy metal systems [40] where the equilibration time (the time needed for the system to transform from a positionally disordered into an ordered state) is of the order of milliseconds, equilibration times for opaline systems could be days and, more likely, centuries or millennia, depending on sustained environmental conditions. If the Debye length is too short the microspherulites will have grown too large to slip past each other easily into the form of a regular array by the time that the Debye shells overlap to give a high electrostatic energy. If the time needed for self-assembly, limited by kinetic factors, is longer than the time available the system will not form the ordered state. Kinematic and, more specifically, molecular dynamical calculations are needed to investigate this matter in detail.

The superlattice constant of the ordered array depends on the number density of nucleation sites initially present. It is also possible that regular opal structures with lattice constants lying outside the specific wave-length range needed to give electromagnetic activity in the visible spectrum (and photonics) do exist, but have not yet been identified in the field because they are not seen to be optically active. Some material identified as potch may be in this category. It should also be kept in mind that we in fact approach the point where control over the conditions required for the laboratory growth of tailored artificial opal will be realised.

Briefly, then, the conditions favourable for the formation of optically active precious opal are (a) a high surface charge density, (b) a long Debye length and (c) an appropriate number density of nucleation centres. A schematic diagram of the proposed process of growth is shown in Fig. 4.

### 3.    Calculation of the energy of interaction between microspherulites

The surface charge density of silica in aqueous solution is anomalously high; much larger than that on the perfect crystalline surface of mica [41], for example, possibly due to the fractal nature of the silica surface which allows charges to be trapped underneath the surface as well as on the outside [29]. Measurements by Bolt [42-43] of the surface charge density of silica in solutions of NaCl of concentration $c$, dissolved in water whose *pH* is made alkaline by the addition of NaOH, are reproduced in Fig. 3. This data suggests a characteristic surface charge density $\sigma \sim -0.1$ C/m$^2$ for alkaline solutions. The distance *D* between the centres of microspherulites is most normally around 250 nm for precious opal [2] and in our model is fixed for a given system by the number of nucleation centres per unit volume.

The electrostatic energies can be high compared to *kT*. For an infinite Debye length *U*/k*T* becomes greater than unity by *R* around 2 nm. A finite Debye length reduces this ratio greatly according to equation (3). The Debye length $L_D$ in equations (2) and (3) depends directly on the electrolyte concentration as given in equation (4). It will always be shorter than 150 nm, the value for de-ionized water in equilibrium with the atmosphere [41]. Some idea of the electrolyte concentration likely to be encountered is obtained from the impurities reported in [3]. It is not, of course, established that all the metallic elements will be in solution, but, for example, if we suppose divalent magnesium to be present at a weight fraction of $7 \times 10^{-3}$ [3] we find a corresponding concentration of $0.3 \times 10^{-3}$ Moles /per litre. And a standard calculation [44] then yields a Debye length ~10 nm.

The dependence of *U*/k*T* as a function of *R* for several Debye lengths is shown in Fig. 5. We assume that, for ordering to take place before contact between the microspherulites at *R* = *D*/2, *U*/k*T* has to be greater than unity. With *T* = 300 K, *D* = 250 nm, a surface charge density of - 0.1 C/m$^2$ and a Debye length of 10 nm it is seen that *U*/k*T* >> 1 for *R* = 60 nm. Under these conditions it may then just be possible for the microspherulites (radius 60 nm, average separation 250 nm) to squeeze past each other to form an ordered array. Longer Debye lengths will make this process easier. Shorter Debye lengths make it harder. Only kinematic calculations can determine this more exactly. Clearly, however, it is to be expected that the system will take a shorter time to self-assemble when *R* is small compared to *D*.

**4.     Calculation of the Debye length**

For the system examined by Bolt [42-43], namely silica in an alkaline NaCl solution, there is an intrinsic competition between the two features needed for the growth of precious opal. Fig. 3 shows that in order to achieve a high surface charge density it is necessary to have a high salt concentration, or a high alkaline *pH*. Both these factors increase the charge density of the electrolyte and this leads to a reduced Debye length, which makes the formation of a regular lattice unfavourable on kinematic grounds. Can these competing factors be reconciled?

We calculate the Debye lengths of the NaCl system considered by Bolt [42-43] in order to determine whether any regime exists for which the formation of precious opal can be considered plausible. The Debye length $L_D$ in an electrolyte [44] is given by:

$$L_D = \sqrt{\frac{\varepsilon_0 \varepsilon_r RT}{2\rho L^2 e^2 I}} = \frac{0.305}{\sqrt{I}} \quad \text{nm} \qquad (4)$$

Where $\rho$ is the density of the solvent, *R* the Gas constant, *L* is Avogadro's number and *I* is the ionic strength:

$$I = \frac{1}{2}\sum_i [i] z_i^2 \qquad (5)$$

[*i*] Being the molar concentration (moles/litre, M/L) and $z_i$ the charge on species *i*.

Since the system consists of NaCl of concentration $c$ dissolved in water whose *pH* is made alkaline by the addition of NaOH it is clear that the modes of chemical dissociation and their corresponding equations of dissociation are:

$$H_2O \Rightarrow H^+ + OH^- \qquad [H^+][OH^-] = K_0 \qquad (6)$$

Where $K_0 = 10^{-14}$ (M/L)$^2$ is the dissociation constant of water at 300 K and the *pH* is given by $[H^+] = 10^{-pH}$

and $\qquad NaOH \Rightarrow Na^+ + OH^- \qquad [Na^+][OH^-] = K_2[NaOH] \qquad (7)$

and $\qquad NaCl \Rightarrow Na^+ + Cl^- \qquad [Cl^-] = c, \qquad (8)$

$K_2$ being the dissociation constant of NaOH, the assumption being made that NaCl is fully dissociated. The equation of electrical neutrality is:

$$[H^+] + [Na^+] = [OH^-] + [Cl^-]. \qquad (9)$$

Since all the ions are monovalent $I$ is given by:

$$I = [H^+] + [Na^+] \qquad (10)$$

From (9) we immediately find that

$$I = K_0/[H^+] + c = 10^{pH-14} + c. \qquad (11)$$

Whence, combining (4) and (11), we can calculate the Debye lengths for the same salt concentrations used by Bolt [42-43] as a function of *pH* (Fig. 6).

It is immediately clear that the only concentrations of NaCl in which the Debye length is longer than 10 nm are the weakest, with $c = 10^{-4}$ and $c = 10^{-3}$. It then follows by examination of Fig. 3 that for the largest surface charge density with these salt concentrations the *pH* must be high – in fact in the interval between 9 and 10. And, accordingly, optically active opal will only be able to form in this system within this narrow range of parameters. It is also the case, of course, that simultaneously the volume density of nucleation centres must be appropriate. This restricted combination of growth conditions itself suggests why natural precious opal is so rare despite the relative high abundance of hydrated silica.

## 5.  Shell structure of the microspherulites

A notable feature of the structure of the individual microspherulites that has been identified by TEM ([3], reproduced in [4]) is the existence of prominent nuclei surrounded by concentric shells. The nuclei have been suggested to contain heavy radioactive elements [4]. We now further suggest that the shells are due to precipitated salt. We assume that the salt solubility limit in the solid phase is much smaller than that in the liquid phase and that the salt concentration in the liquid phase is below its solubility limit. As the growth of the solid microspherulites progresses, the salt that is in the volume of the solid phase is expelled into the liquid. Accordingly, the salt concentration ahead of the solidification front grows continuously larger. At some stage it reaches the solubility limit of the liquid and the salt rapidly precipitates onto the surface of the microspherulite. This produces a shell of the most abundant solid salt in the liquid at that point. The process is essentially a means of separating the salts out radially – it can repeat itself larger radii – thereby providing for a multiple spheroidal shell structure. A proper microchemical analysis of the shell structure would resolve this.

## 6.    Conclusions

We have outlined a basic model of the growth conditions of ordered precious opal based upon electrostatic self-assembly. Conditions favourable for growth are those which produce a high surface charge and a long Debye length. In addition, the number density of nucleation sites must be appropriate. This restricted combination of growth conditions does itself, by implication, suggest why precious opal is so rare in nature. For silica growing in a sodium chloride solution a high alkaline *pH* and a low salt concentration are necessary conditions. In post-nucleation growth of individual microspherulites it is proposed that sequential deposition of abundant intrinsic salts is responsible for a concentric spherical shell structure about central nuclei. Sedimentation of spherulites under gravity is a quite separate physical process which may take place later.

**Appendix A**

We consider a solid spherulite of radius $R$ which is growing in an infinitely large supersaturated solution as tiny moieties diffuse from the outer reaches of the solution towards the spherulite, finally condensing on its surface. The driving force for diffusion is the gradient of the chemical potential $\mu(r)$ of the moieties in the liquid at a distance $r$ ($r \geq R$) from the centre of the spherulite. The number flux density $\boldsymbol{J}(r)$ of flow of the moieties is given by Fick's law:

$$\boldsymbol{J}(\boldsymbol{r}) = -D\nabla\mu(\boldsymbol{r}) \tag{A1}$$

where $D$ is a diffusion constant. The steady state is characterised by $\nabla^2\mu(\boldsymbol{r}) = 0$ whose solution is [45]:

$$\mu(r) = \mu(\infty) - \Delta\mu\frac{R}{r} \tag{A2}$$

where $\Delta\mu = \mu(\infty) - \mu(R)$. The quantity $\mu(\infty)$ is the chemical potential of the moieties in the unperturbed solution and $\mu(R)$ is the chemical potential at the surface of a spherulite which, assuming quasi-equilibrium and the absence of surface effects, is equal to the chemical potential in the solid and so is independent of $R$.

The gradient of the chemical potential at $r = R$ is $\Delta\mu/R$ and the diffusive flux density is $-D\Delta\mu/R$. The rate of increase of the volume of the spherulite is given by multiplying the magnitude of this by $4\pi r^2 v$, where $v$ is the volume of a moiety and so:

$$\frac{\partial R}{\partial t} = \frac{vD\Delta\mu}{R} \tag{A3}$$

or $$t = \frac{1}{vD\Delta\mu}\int_{R_0}^{R} R'\,\mathrm{d}R' \tag{A4}$$

where $R_0$ is the initial radius of the nucleation centre. When integrated, this gives $R = \sqrt{R_0^2 + 2vD\Delta\mu t}$, showing that at long times the radius of the spherulite increases with the square root of the time. As a result, if a second spherulite starts to grow at a time $\tau$ after an earlier first, then the relative radii $r$, areas $a$ and volumes $V$ at large time will proceed as $\Delta r/r = \tau/2t$, $\Delta a/a = \tau/t$, $\Delta V/V = 3\tau/2t$ respectively, and the spherulites become effectively equal in size after a time $t \gg \tau$. It is of course accepted that, during the process of growth, $\Delta\mu$ may change with alterations in the environmental conditions, for example the periodic daily or annual variations of temperature and rainfall.

**Acknowledgements**
Support from the CSIRO (Australia) is acknowledged by one of us (LTC), as is the generous hospitality of the Master and Fellows, Gonville and Caius College, Cambridge.

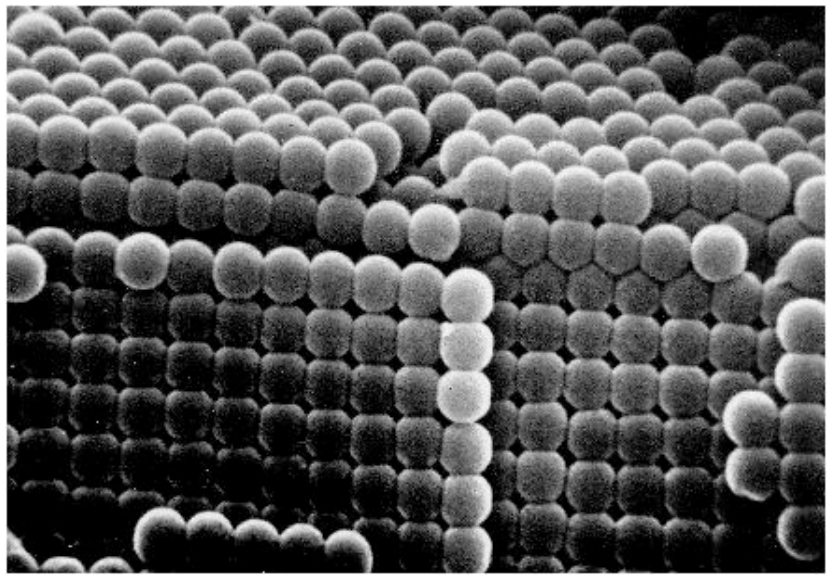

Fig. 1.	Scanning electron micrograph of a sliver of precious opal from a mine around Dubnik, Slovakia, taken by Professor Max Weibel and co-workers at the Swiss Institute of Technology in Zurich [1]. The spherulites are ~ 0.5 μm in diameter and are situated on a periodic lattice.

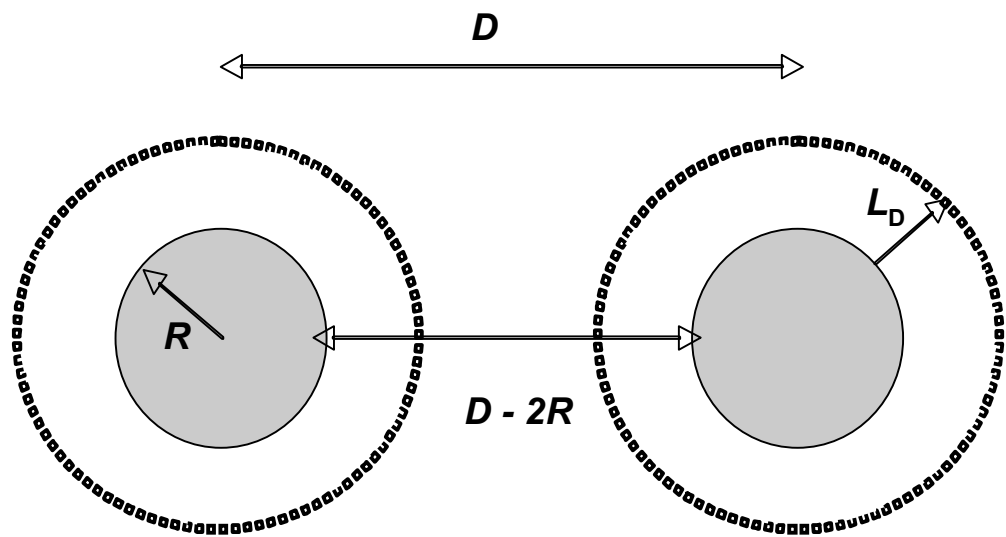

Fig. 2.    Schematic diagram of two solid microspherulites of hydrated silica (shaded) surrounded by their Debye shells (dashed circles).

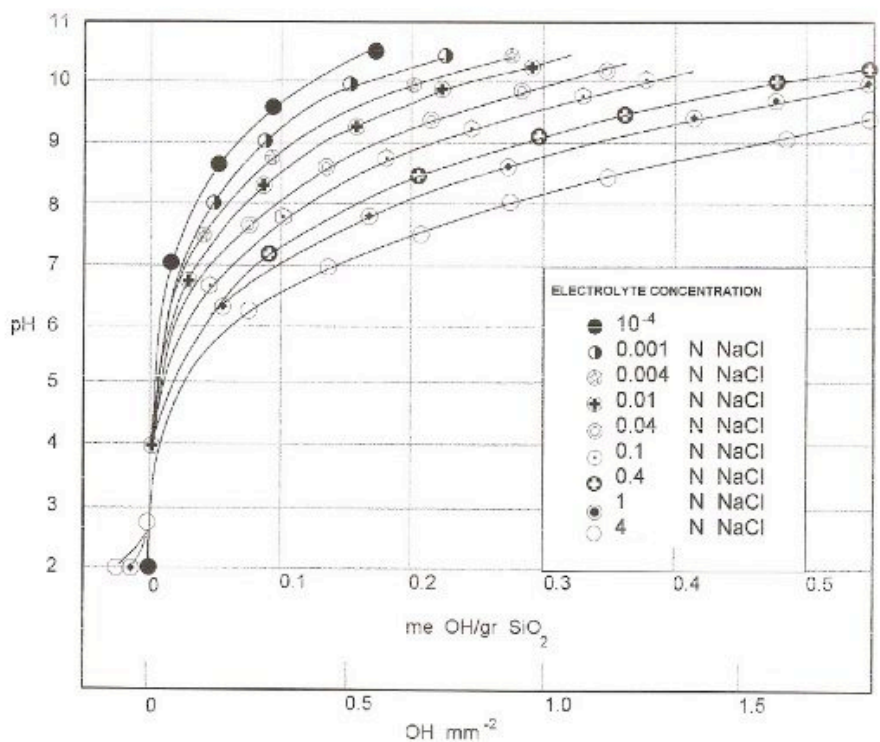

Fig. 3.    Data of Bolt [42-43] for the surface charge density on colloidal silica surfaces with a specific surface area of 180 $m^2 g^{-2}$ in an aqueous solution of NaCl as a function of *pH* for various concentrations of NaCl. The *pH* is varied by adding NaOH. One electron per $nm^2$ is equivalent to a charge density of 0.16 $C/m^2$. (Courtesy of the Editors, Journal of Physical Chemistry)

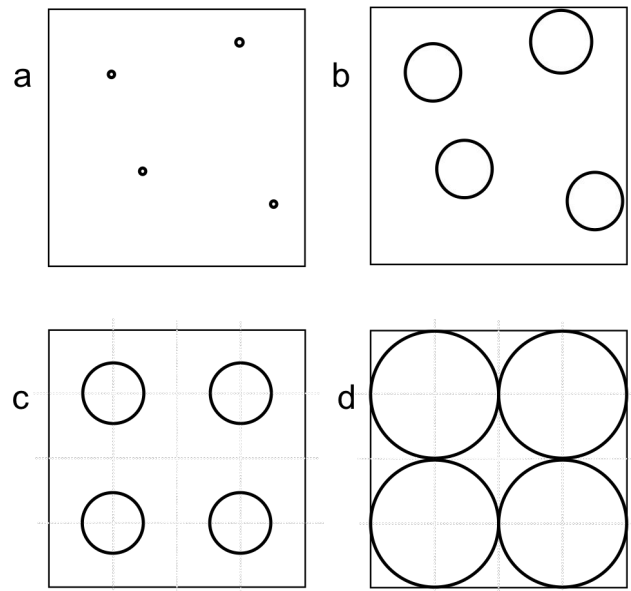

Fig. 4. Schematic diagram of the proposed growth mechanism for precious opal. (a) Growth of the opal spherulites starts at nucleation centres positioned randomly in the solution. Four of them are shown here. When the spherulites reach a critical size (b) electrostatic forces overwhelm entropic effects causing the spherulites to line up in an ordered array (c). Growth continues until the solute is exhausted (d). The distance between the centers of the spherulites in optically active opal corresponds to the wavelength of visible light of a few hundred nm.

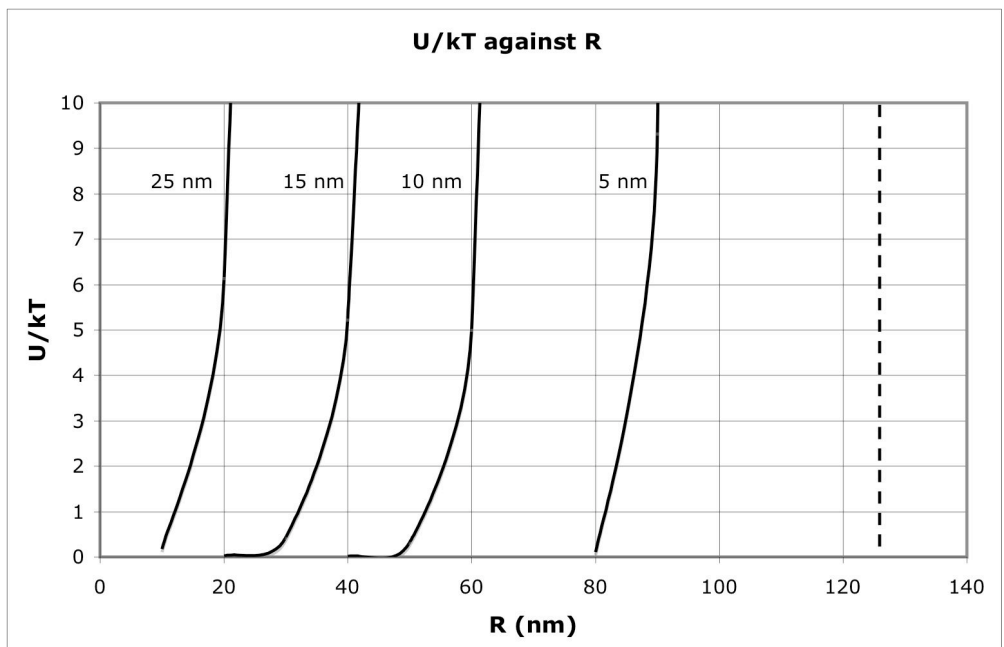

Fig. 5. Ratio of electrostatic potential energy $U$ of interaction to $kT$ as a function of $R$ between two charged spheres of radii $R$ whose centres are separated by a distance $D$, for given surface charge density $\sigma$ for several Debye lengths, following equation (3). Values of the parameters used in the calculation are $D = 250$ nm, $\sigma = -0.1$ C/m$^2$, $\varepsilon_0 = 8.85 \times 10^{-12}$ C$^2$/Nm$^2$, $\varepsilon_r = 80$. The Debye lengths used are 5 nm, 10 nm, 15 nm, 25 nm. The vertical dashed line shows the value of $R$ (125 nm) at which the two spheres make contact.

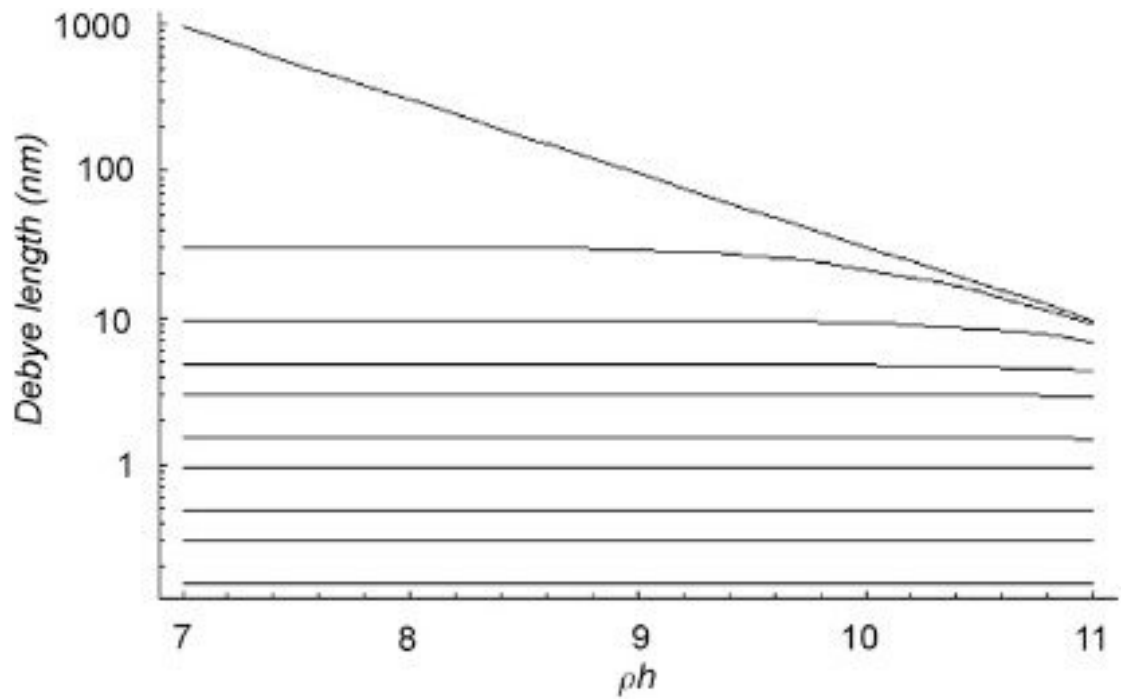

Fig. 6.         Calculated Debye lengths at a temperature of 300 K against *pH*, following equation (11) for the different concentrations *c* of NaCl in water used by Bolt [42-43]. The top curve is for pure water, the next one down for $c = .0001$ etc. in order as $\{c = 0, 0.0001, 0.001, 0.004, 0.01, 0.04, 0.1, 0.4, 1, 4\}$.